\documentclass{aa}  

\usepackage{graphicx}
\usepackage{color}
\usepackage{amssymb}

\begin{document} 

\def\arcsec{\hbox{$^{\prime\prime}$}}

   \title{Detection of H$_2$O and OH$^+$ in $z>3$ Hot Dust-Obscured Galaxies}
   \titlerunning{H$_2$O and OH$^+$ in $z>3$ Hot DOGs}
   \authorrunning{F. Stanley et al.}

   \author{F. Stanley,\inst{1}\thanks{flora.stanley@chalmers.se}
          K. K. Knudsen,\inst{1}
          S. Aalto,\inst{1}
          L. Fan,\inst{2,3}
          N. Falstad,\inst{1}
          E. Humphreys\inst{4,5}
          }

   \institute{Department of Space, Earth and Environment, Chalmers University of Technology, Onsala Space Observatory, SE-439 92 Onsala, Sweden 
   \and  CAS Key Laboratory for Research in Galaxies and Cosmology, Department of Astronomy, University of Science and Technology of China, Hefei 230026
   \and School of Astronomy and Space Sciences, University of Science and Technology of China, Hefei, Anhui 230026, People's Republic of China 
   \and Joint ALMA Observatory (JAO) Alonso de C\'ordova 3107, Santiago, Chile  
   \and European Southern Observatory, Alonso de C\'ordova 3107, Casilla 19001, Vitacura, Santiago, Chile                   
             }


 
  \abstract
  {} 
   {In this paper we present the detection of H$_{2}$O and OH$^+$ emission in $z>3$ hot dust-obscured galaxies (Hot DOGs).}
   {Using ALMA Band-6 observations of two Hot DOGs, we have detected H$_{2}$O($2_{02}-1_{11}$) in W0149$+$2350, and H$_{2}$O($3_{12}-3_{03}$) and the multiplet OH$^+$($1_{1}-0_{1}$) in W0410$-$0913. These detections were serendipitous, falling within the side-bands of Band-6 observations aimed to study CO(9$-$8) in these Hot DOGs.}
   {We find that both sources have luminous H$_{2}$O emission with line luminosities of $L_{\rm H_2O(2_{02}-1_{11})} > 2.2\times10^8 L_\odot$ and $L_{\rm H_2O(3_{12}-3_{03})} = 8.7\times10^8 L_\odot$ for W0149$+$2350 and W0410$-$0913, respectively. The H$_{2}$O line profiles are similar to those seen for the neighbouring CO(9$-$8) line, with linewidths of FWHM $\sim 800-1000$\,km\,s$^{-1}$. However, the H$_{2}$O
   	   emission seems to be more compact than the CO(9$-$8).
   OH$^+$($1_{1}-0_{1}$) is detected in emission for W0410$-$0913, with a FWHM$=1000$\,km\,s$^{-1}$ and a line luminosity of $L_{\rm OH^+(1_{1}-0_{1})} = 6.92\times10^8 L_\odot$. The ratio of the observed H$_{2}$O line luminosity over the IR luminosity, for both Hot DOGs, is consistent with previously observed star forming galaxies and AGN. The H$_2$O/CO line ratio of both Hot DOGs and the OH$^+$/H$_2$O line ratio of   W0410$-$0913 are comparable to those of luminous AGN found in the literature.}
   {The bright H$_2$O($2_{02}-1_{11}$), and H$_2$O($3_{12}-3_{03}$) emission lines are likely due to the combined high star-formation levels and luminous AGN in these sources. The presence of OH$^+$ in emission, and the agreement of the observed line ratios of the Hot DOGs with luminous AGN in the literature, would suggest that the AGN emission is dominating the radiative output of these galaxies. However, follow-up multi-transition observations are needed to better constrain the properties of these systems.}


   \maketitle
%
\section{Introduction}
Energetic feedback from active galactic nuclei (AGN) has the potential to heat and remove 
the cold molecular gas from the interstellar medium (ISM), consequently suppressing the star-formation 
of the host galaxy. The most successful models of galaxy formation and evolution require AGN feedback 
to explain many of the puzzling properties of local massive galaxies and the intergalactic medium 
(e.g., red colours, steep luminosity function, black-hole – spheroid relationship, metal enrichment of the intergalactic medium; 
see \citealt{AlexanderHickox12}, and \citealt{Fabian12} for reviews).
Understanding the properties of the molecular gas in the ISM of active galaxies is key to constraining and 
disentangling the effects of AGN and star formation. 

Within the past few years the Wide-Infrared Survey Explorer \citep[WISE;][]{Wright10} has led to the discovery of a 
population of luminous, dust-obscured AGN \citep[e.g.][]{Eisenhardt12,Bridge13,Lonsdale15}. 
Selected to be the reddest sources in the WISE colour-colour plot ([3.4-4.6]\,$\mu$m vs [4.6-12]\,$\mu$m), 
they are by definition highly obscured and dusty galaxies.
 They are commonly referred to as Hot Dust Obscured Galaxies 
(Hot DOGs). This new class of galaxies is a rare sample with only 1000 over the entire
sky \citep[e.g.,][]{Eisenhardt12}. 
They are mostly high-redshift objects, at $1<z<4$ \citep[e.g.][]{Wu12, Tsai15}, and extremely luminous with 
$L_{bol} > 10^{13}L_\odot$ \citep[e.g.][]{Jones14,Wu14}. Based on X-ray observations and spectral energy distribution (SED) studies, there is clear evidence that they host highly dust-obscured AGN \citep[e.g.][]{Stern14,Piconcelli15,Assef15,Assef16,Fan16b}.
 
Hot DOGs are found in predominantly over-dense regions \citep[e.g.][]{Jones17}, 
and the most luminous among them show  evidence of a high merger fraction \citep[62\%][]{Fan16a}. 
This supports a scenario where Hot DOGs 
could represent the post-merger transitional phase from a dusty starburst 
dominated phase to an optically bright quasar phase \citep[e.g.,][]{Eisenhardt12,Wu12,Fan16a}.  
Their rarity could be a result of rare
 events such as major mergers, the relative briefness of the transitional 
phase they represent, and/or the population is tracing the tail end of the mass 
 function.  

In this paper we present the detection of H$_2$O, and OH$^+$ emission lines for the first time in Hot DOGs. 
In the past few years there have been a number of detections of H$_2$O transitions in the far-infrared (FIR) and 
sub-mm wavelengths with the advent of observatories such as {\it Herschel}, NOEMA, and ALMA. 
Studies of nearby galaxies have shown that H$_2$O is the third most abundant species in dense warm star-forming 
regions, and shock heated regions \citep[e.g.][]{Cernicharo06, Bergin03, Gonzalez-Alfonso13}. 
Detections of H$_2$O transitions in high-$z$ and lensed sub-mm galaxies confirm that they are among the strongest molecular 
lines in starburst galaxies and also show similar line profiles and spatial distribution as the high-$J$ CO emission 
\citep[e.g.][]{Lis11,vanderWerf11,Bradford11,Combes12,Lupu12,Bothwell13,Omont13,Riechers13,Weiss13,Yang16,Oteo17,Yang19,Apostolovski19, Casey19,Jarugula19,Yang20}.  
Furthermore, there is a correlation between the line luminosity and the IR luminosity, extending over three orders of magnitude 
\citep[e.g.][]{Omont13, Yang13, Yang16}. Modelling of H$_2$O emission in extragalactic sources has shown that the combination 
of different transitions can help disentangle the different components of the molecular ISM \citep[e.g.][]{Gonzalez-Alfonso14, Liu17}.

OH$^+$ is an important part of the chemical network for the formation of H$_2$O, and can trace the ionisation rate of 
atomic/molecular gas \citep[e.g.][]{Neufeld10,Hollenbach12}, as well as the turbulent gas component \citep[e.g.][]{Gonzalez-Alfonso18}. 
OH$^+$ is observed both in emission and absorption in both low and high-$z$ galaxies \citep[e.g.][]{vanderWerf10, Spinoglio12, 
Kamenetzky12, Pereira-Santaella13, Riechers13, Gallerani14, vanderTak16, Li20}. 
Interestingly, OH$^+$ is observed in emission for sources with a strong AGN component, \citep[e.g.][]{vanderWerf10,Spinoglio12,Pereira-Santaella13,Li20}, while starburst galaxies show OH$^+$ in absorption \citep[e.g.][]{Kamenetzky12, Riechers13, vanderTak16}.

The two Hot DOGs discussed in this paper, W0149+2350 and 
W0410-0913, were selected to be among the most starbursting Hot DOGs, with massive 
molecular gas reservoirs discovered in CO(4$-$3) (Fan et al. 2018).
Both sources have exceptionally wide CO lines in multiple transitions, 
with FWHM$ \approx 750-950$\,km\,s$^{-1}$ \citep[][Knudsen et al. in prep; see Table~1]{Fan18} 
and large molecular gas masses of $10^{10} - 10^{11} M_\odot$. 

In section 2 we present the ALMA observations and the extraction of the spectra. 
In section 3 we present our results for the two sources, and in section 4 
we discuss the cause of excitation of the observed emission and compare it to the literature.
Finally, in section 5 we give a summary and conclusions.
Throughout this paper we assume a WMAP7 cosmology.

\section{ALMA observations and analysis}
We obtained multi-band observations of W0149$+$2350 and W0410$-$0913 in project
2017.1.00123.S to study CO rotational transitions, and these data yielded
three serendipitous detections of other molecular lines. 
Observations were carried out using the band-6 receivers, where one side-band
was tuned to the redshifted CO(9-8) with the correlator used in frequency
domain mode with each spectral window (spw) having a bandwidth of 1.875\,GHz,  
the other sideband used a continuum setup. Observations of W0149$+$2350 were carried 
out on 24th of September 2018, with a total of 49\,min of on-source integration time. 
The observed frequencies cover the ranges of 229.82--233.60\,GHz, and 243.68--245.55\,GHz, 
which include the redshifted H$_{2}$O($2_{02}-1_{11}$) transition in addition 
to the targeted CO(9-8). Observations of W0410$-$0913 were carried out on the 07th of September 2018, 
with a total of 6\,min of on-source integration time. The observed frequencies cover the ranges of 
221.08--224.81\,GHz, and 236.28--240.57\,GHz, which include the H$_{2}$O($3_{12}-3_{03}$) transition, 
and the OH$^+$($1_1-0_1$) multiplet in addition to the targeted CO(9-8). Flux and bandpass 
calibration was done using J0006$-$0623 and J0423$-$0120, and gain calibration was done with 
J0152+2207 and J0407$-$1211, for W0149$+$2350 and W0410$-$0913, respectively. Beam sizes correspond to $0.53\arcsec\times0.32\arcsec$, and $0.67\arcsec\times0.52\arcsec$, for W0149$+$2350 and W0410$-$0913 respectively.

Reduction, calibration, and imaging was done using {\sc CASA} (Common
Astronomy Software Application\footnote{https://casa.nrao.edu};
\citealt{McMullin07}).   
The pipeline reduced data delivered from the observatory were of sufficient
quality, no or little additional flagging and further calibration was necessary.  
The steps required for standard reduction and
calibration included in the pipeline include flagging, bandpass calibration, as well
as flux and gain calibration. For our imaging process we applied natural weighting.
A conservative estimate of the error on the absolute flux calibration is $10\%$. 
Continuum emission was subtracted using the line-free channels identified in
both sidebands for each source.
For both sources, the channel width was set to 60\,km\,s$^{-1}$. 

To extract the spectrum we used the following procedure. 
We created the moment-0 map of each emission line, and used the {\sc imfit} tool of {\sc casa} to get an 
estimate of the extent of the emission. We then extracted the spectrum from a region that is equivalent 
to two times the size estimate of the source.
 
In this paper we also show the CO(9$-$8) emission for comparison. This data has been extracted and analysed in the same way as the H$_{2}$O and OH$^+$ emission that are the focus of this paper. However, the CO(9$-$8) data will be presented in more detail in Knudsen et al. (in prep) along with other CO transitions. 

\begin{table*}
	\centering
	\caption{Source properties as presented in \cite{Fan18}. }\label{tab:sources}
	\begin{tabular}{lccccccccc}
\hline 
\hline 

Source & RA & DEC & $z_{\rm CO(4-3)}$ & $L'_{\rm CO(4-3)}$ & $L_{\rm IR,AGN}$ & $L_{\rm IR,SF}$ \\
&  [J2000] & [J2000] & & [K\,km\,s$^{-1}$\,pc$^2$] & [$L_\odot$] & [$L_\odot$]  \\ 
\hline \noalign {\smallskip}
W0149$+$2350 & 01:49:46.16 & $+$23:50:14.6 & 3.237 & $2.4 \times 10^{10}$ & $7.6\times10^{13}$ & $1.14\times10^{13}$ \\
W0410$-$0913 & 04:10:10.60 & $-$09:13:05.2 & 3.631 & $1.8 \times 10^{11}$ & $1.5 \times 10^{14}$ & $4.7\times10^{13}$   \\
\hline
		\end{tabular}
\end{table*}

\section{Results}

\begin{table}
	\centering
	\caption{Identified lines based on the CDMS\protect\footnotemark catalogue entries, and corresponding RMS of their spectrum (RMS$_{\rm spec}$).}\label{tab:lines}
	\begin{tabular}{lcc}
\hline 
\hline 
Line & $\nu_{\rm rest}$ & RMS$_{\rm spec}$ \\
& [GHz] & [mJy] \\
\hline
H$_2$O($2_{02}-1_{11}$) & 987.937 & 0.28\\
H$_2$O($3_{12}-3_{03}$) & 1097.365 & 0.30\\
OH$^+$($1_{1}-0_{1}$) multiplet: & & 0.57 \\	
OH$^+$($1_{11}-0_{11}$) & 1032.998 \\
OH$^+$($1_{12}-0_{11}$) & 1033.004 \\
OH$^+$($1_{11}-0_{12}$) & 1033.113 \\
OH$^+$($1_{12}-0_{12}$) & 1033.118 \\ 
\hline
\end{tabular}
\end{table}

\subsection{H$_2$O(2$_{02}$--1$_{11}$) in W0149$+$2350}
\begin{figure}
\centering
\includegraphics[width=0.45\textwidth]{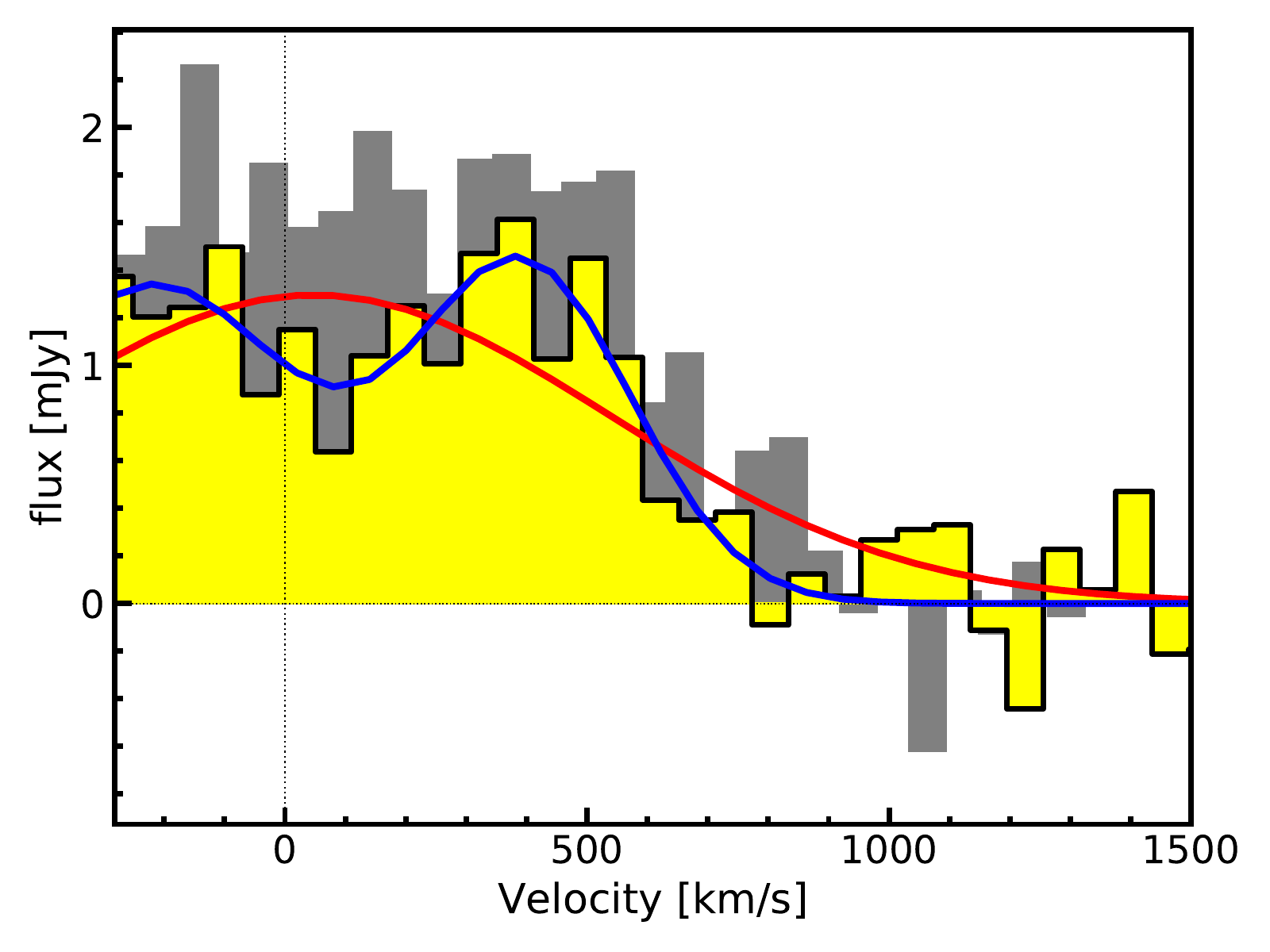} 
\caption{The detected H$_2$O line (yellow) of W0149$+$2350, with the CO($9-8$) emission (grey) shown in the background for comparison. The fitted single and double Gaussians are shown with the red and blue curves respectively. 
		}\label{fig:specfit_w0149}
\end{figure}

For W0149$+$2350 we have detected H$_2$O emission in the transition of H$_2$O(2$_{02}$--1$_{11}$), see Fig.~\ref{fig:specfit_w0149} (in yellow). Although
the line is truncated, it covers $\rm \sim1000\,km\,s^{-1}$ with a line profile that cannot be described by a single Gaussian.
This type of line profile is also seen in their CO(9-8) line emission, see Fig.~\ref{fig:specfit_w0149} (in grey; Knudsen et al. in prep). 

In Fig.~\ref{fig:specfit_w0149} we show the line, with the fit of a single Gaussian, and a double Gaussian.
A double Gaussian fit to the line improves the $\chi^2$ by 22\% compared to the single Gaussian fit. 
In Table~\ref{tab:results} we give the fit results for the two component fit to the line. However, as the line is truncated, we integrate the measured flux of the line, to calculate a lower limit of the total integrated line flux (see Table~\ref{tab:results}). 

\begin{figure}
\centering
\includegraphics[width=0.45\textwidth]{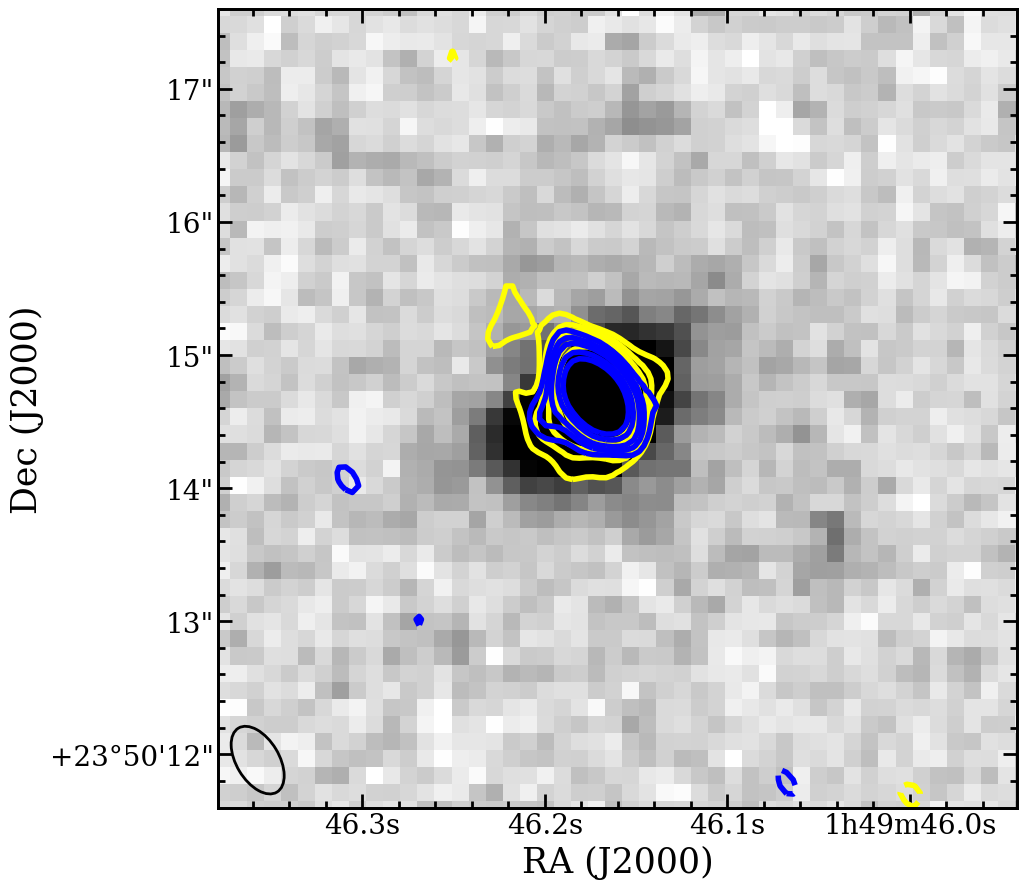}\\	
\caption{HST WFC3 cutout of W0149+2350.  Overplotted are the contours corresponding to 3, 4, 5, 7, and 9$\sigma$, for the H$_2$O line (blue), and CO($9-8$) line (yellow). The RMS values of the moment maps correspond to 0.046\,Jy\,km\,s$^{-1}$ and 0.055\,Jy\,km\,s$^{-1}$ for the H$_2$O and CO($9-8$) lines, respectively. Negative contours corresponding to -3$\sigma$ are also plotted with dashed lines.}\label{fig:hst_w0149}
\end{figure}

In Fig.~\ref{fig:hst_w0149} we show the HST map from \cite{Fan16a}, with the moment-0 map contours of the H$_2$O emission in blue, and the CO contours in yellow for comparison.
To estimate the extent of the emission observed, we use the {\sc imfit} tool of {\sc casa}, to fit to the moment-0 maps of each line. However, as the H$_2$O line is truncated, to have a meaningful comparison of the extent of the emission we create moment-0 maps limited to the velocity range of $-300 < \upsilon < 750$\,km\,s$^{-1}$, for both the H$_2$O and CO emission. We find that the H$_2$O emission 
of W0149$+$2350 extends across $2.3\pm0.6$\,kpc, while the CO(9$-$8) emission extends across $3.3\pm0.5$\,kpc (see Table~\ref{tab:size}).

\footnotetext{https://cdms.astro.uni-koeln.de}

\begin{table*}
	\centering
	\caption{Detected line properties. For each emission line we give the central velocity ($\upsilon_{\rm cen}$), peak flux ($S_{\rm peak}$), FWHM, and integrated flux ($S_{\rm int}$), based on the Gaussian fits to the line. We also give the corresponding line luminosity ($L_{\rm line}$). The H$_2$O emission for both sources is best fit by a double Gaussian, therefore we give the parameters of the two components corresponding to the fit. In the case of W0149$+$2350, where the line is truncated, we give a lower limit on the total $S_{\rm int}$ and $L_{\rm line}$ based on the integration of the observed spectral line.  }\label{tab:results}
	\begin{tabular}{llcccccccc}
\hline 
\hline 

Source & line & $\upsilon_{\rm cen}$ & $S_{\rm peak}$ & FWHM & $S_{\rm int}$  & $L_{\rm line}$ \\ 
&   & [km\,s$^{-1}$] & [mJy] & [km\,s$^{-1}$] & [Jy\,km\,s$^{-1}$]  & [$\times10^{8}L_\odot$] \\ 
\hline \noalign {\smallskip}
W0149$+$2350 &  H$_2$O($2_{02}-1_{11}$) \\
& comp1 & $-219\pm103$ & $1.3\pm0.2$ & $564\pm413$ & $0.80\pm0.60$ & $1.58\pm1.17$\\
& comp2 & $396\pm58$ & $1.4\pm0.2$ & $421\pm99$ & $0.63\pm0.18$ & $1.24\pm0.36 $\\
& {\bf total} & & & &  $>$1.12 & $>2.2$ \\
\hline \noalign {\smallskip}
W0410$-$0913 & H$_2$O($3_{12}-3_{03}$) \\
& comp1 & $-307\pm44$ & $1.2\pm0.7$ & $256\pm167$ & $0.31 \pm 0.29$ & $0.83\pm0.76$ \\
& comp2 & $99\pm54$ & $3.8\pm0.2$ & $737\pm103$ & $2.99\pm0.46$ & $7.90\pm1.23$ \\
& {\bf total} & & & & $3.30\pm0.54$ & $8.73\pm 1.44$ \\ [6pt]
 & OH$^+$($1_{1}-0_{1}$) &  $-100\pm83$   & $2.6\pm0.3$ & $1000\pm196$ & $2.78\pm0.62$ & $6.92\pm1.55$ \\
\hline
		\end{tabular}

\end{table*}

\subsection{H$_2$O(3$_{12}$--3$_{03}$) and OH$^+$($1_1-0_1$) in W0410$-$0913}

For W0410$-$0913 we have detected the H$_2$O($3_{12} - 3_{03}$), and the OH$^+$($1_1-0_1$) multiplet with transitions at 1032.998--1033.118\,GHz (see Table~\ref{tab:lines}, although unresolved). 
Similar to what is seen for W0149$+$2350, the H$_2$O($3_{12} - 3_{03}$) line also shows a very wide profile, that is similar to that of the CO(9$-$8). 

\begin{figure}
\centering
\includegraphics[width=0.45\textwidth]{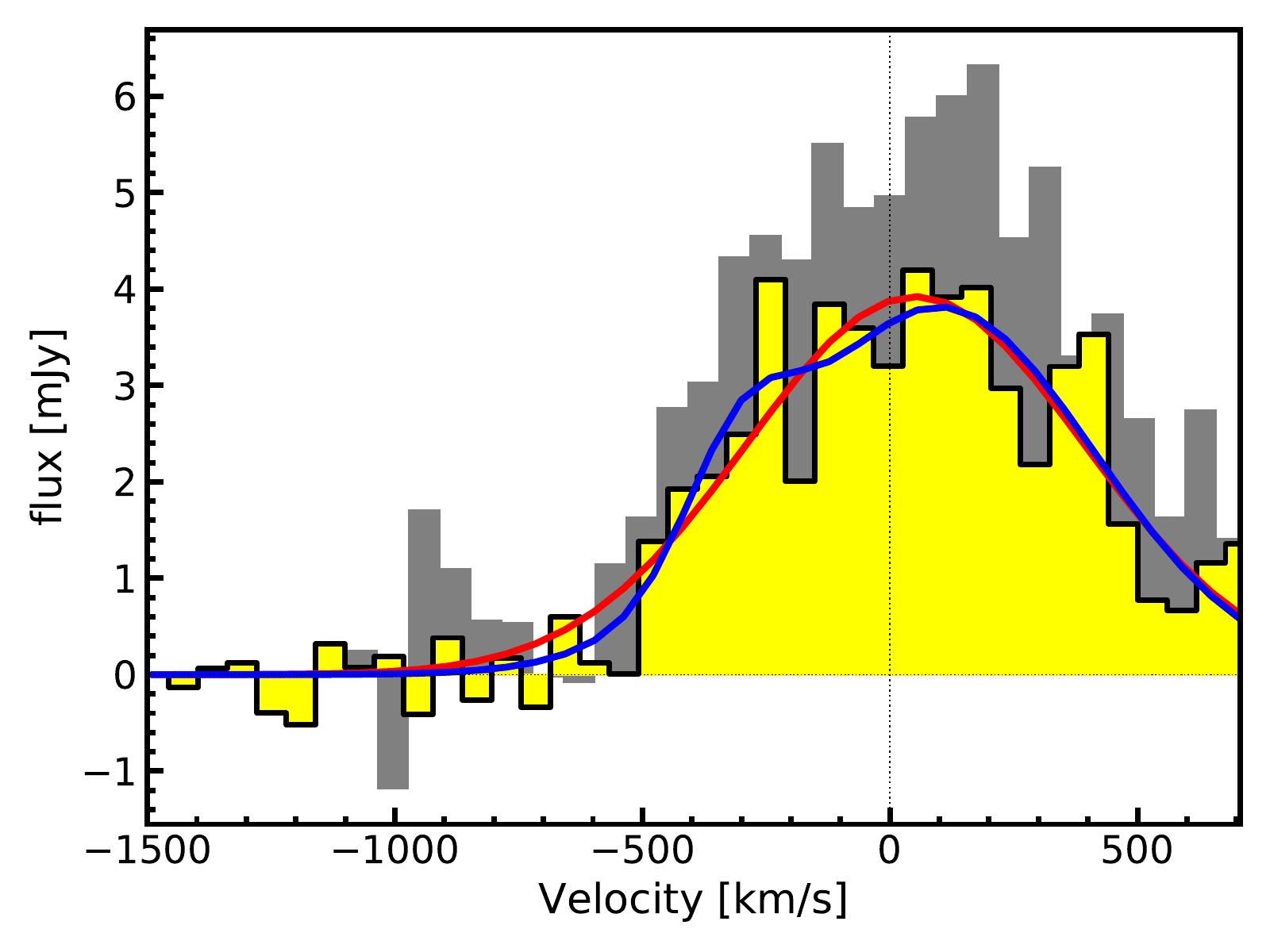} 
\caption{The detected H$_2$O line (yellow) of W0410$-$0913, with the CO($9-8$) emission (grey) shown in the background for comparison. The fitted single and double Gaussians are shown with the red and blue curves respectively.  
		}\label{fig:specfit_w0410}
\end{figure}

In Fig.~\ref{fig:specfit_w0410} we show the H$_2$O($3_{12} - 3_{03}$) line, with the fit of a single Gaussian, and a double Gaussian. A double Gaussian fit to the line improves the $\chi^2$ by 12\% compared to the single Gaussian fit. 
In Table~\ref{tab:results} we give the results for the two component fit to the line.

\begin{figure}
\centering
\includegraphics[width=0.45\textwidth]{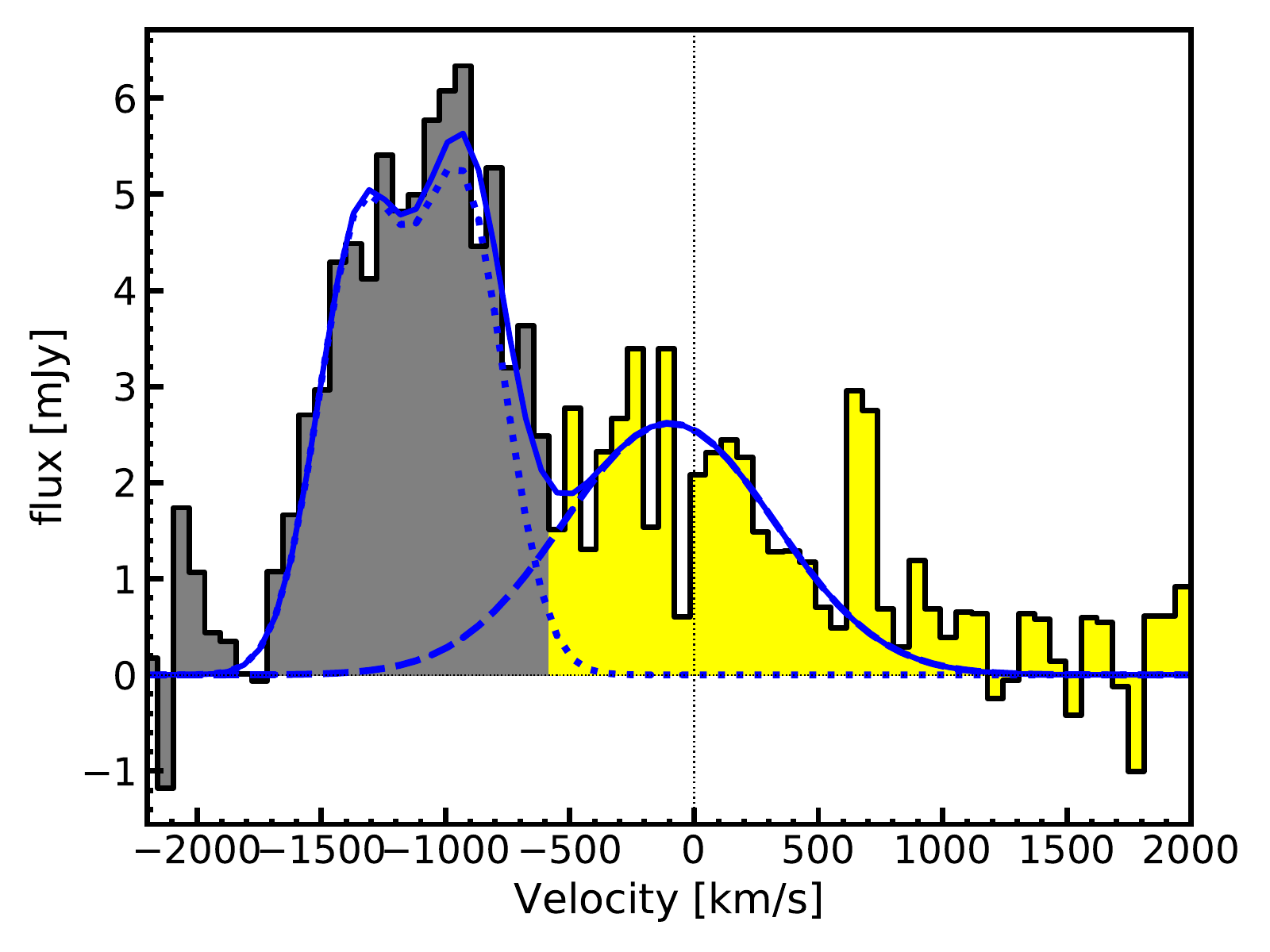} 
\caption{The detected OH$^+$ line (yellow) of W0410$-$0913, next to the CO(9$-$8) line (grey). The grey colour also shows the channels used to fit the CO(9$-$8) line. The blue dotted curve shows the double Gaussian fit to the CO($9-8$), the blue dashed curve shows the single Gaussian fit to the OH$^+$ emission, and the solid curve is the combination of the above.}\label{fig:specfitoh_w0410}
\end{figure}

Due to their large linewidths the CO(9$-$8) and OH$^+$ lines are partially blended. 
We perform a simultaneous fit of the CO(9$-$8) and OH$^+$ emission. We do this by using a double Gaussian fit for the CO(9$-$8), and a single Gaussian for the OH$^+$ emission. We allow all three parameters of the Gaussians to vary during the fit, restricting the central velocity ($\upsilon_{\rm cen}$) of each component of the CO line within a range of 400\,km\,s$^{-1}$, and for the OH$^+$ line within 200\,km\,s$^{-1}$ of the expected $\upsilon_{\rm cen}$.
In Fig.~\ref{fig:specfitoh_w0410} we show the double Gaussian fit to the CO(9$-$8) line (blue dotted curve), and the resulting fit to the OH$^+$ emission (blue dashed curve). In Table~\ref{tab:results} we give line properties based on the fit. In this case the integrated line flux ($S_{\rm int}$) is based on the fit.
We note that in our analysis we have assumed that the OH$^+$ is in emission based on the observed spectrum; however, it is possible that it has a different line profile, such as a P-Cygni profile, that is not visible due to the blending with the CO(9$-$8) line. In some starburst galaxies OH$^+$ has been found to have a P-Cygni line profile. An example is NGC253, for which it is concluded that the OH$^+$ is primarily arising in cold diffuse foreground gas \citep{vanderTak16}. If indeed the OH$^+$ multiplet observed here had a P-Cygni profile, it would require a strongly asymmetric line profile for the CO(9-8) in order to compensate for the superimposed absorption feature of the OH$^+$. This would require luminous CO(9$-$8) emission at $\sim$1000\,km\,s$^{-1}$ from the line centre. Such a peculiar line profile is unlikely, and is not seen for the lower$-J$ CO \citep[][Knudsen et al. in prep]{Fan18} and H$_2$O transitions. 
It could also be possible for the emission identified as OH$^+$ to be alternatively interpreted as a massive and highly excited molecular outflow. However, such a scenario would require a combination of orientation and foreground absorption that would allow us to see the red component of the outflow while not seeing emission from the blue component. Furthermore, there is no evidence for such a massive outflow in the lower$-J$ CO transitions \citep[][Knudsen et al. in prep]{Fan18}.

\begin{figure}
\centering 
\includegraphics[width=0.45\textwidth]{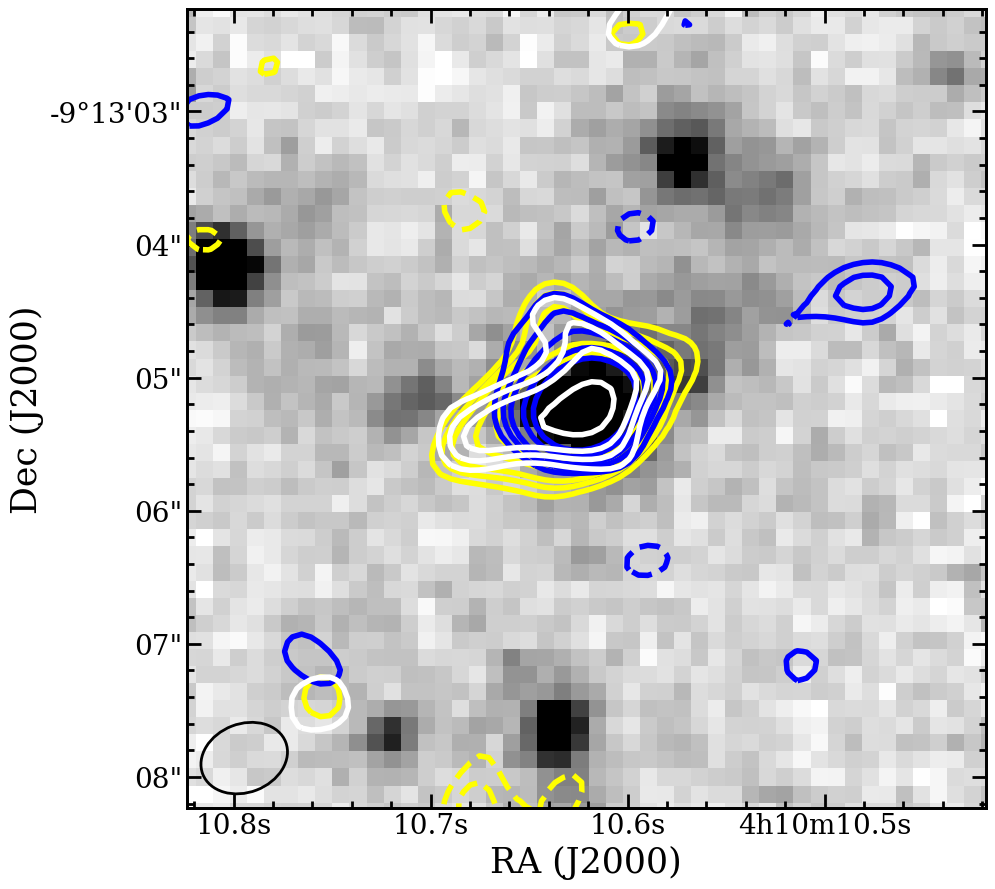} \\	
\caption{HST WFC3 cutout of W0410-0913. Overplotted are the contours corresponding to 3, 4, 5, 7, and 9$\sigma$, for the H$_2$O line (blue), OH$^+$ line (white), and CO($9-8$) line (yellow). The RMS values of the moment maps correspond to 0.12\,Jy\,km\,s$^{-1}$, 0.146\,Jy\,km\,s$^{-1}$, and 0.116\,Jy\,km\,s$^{-1}$ for the H$_2$O, OH$^+$ and CO($9-8$) lines, respectively. Negative contours corresponding to -3$\sigma$ are also plotted with dashed lines.}\label{fig:hst_w0410}
\end{figure}

In Fig.~\ref{fig:hst_w0410} we show the HST map from \cite{Fan16a}, with the moment-0 map contours of the H$_2$O emission in blue, the OH$^+$ emission in white, 
and the CO contours in yellow for comparison. The OH$^+$ emission follows closely the extent seen in the CO(9$-$8), that shows an extended non-uniform structure, but the 
H$_2$O emission is more compact. 
We estimate the extent of the emission observed using the {\sc imfit} tool of {\sc casa}. Following the analysis of W0149$+$2350, we create moment-0 maps limited to the velocity range of $-500 < \upsilon < 500$\,km\,s$^{-1}$, for the H$_2$O, CO, and OH$^+$ emission, to match to the velocity range covered by the H$_2$O line. The OH$^+$ is extended across $9.3\pm2.1$\,kpc, similar to the CO($9-8$) emission extending over $5.7\pm0.6$\,kpc, while the H$_2$O is more compact covering $3.0\pm0.6$\,kpc. The extent of the CO(9$-$8) and OH$^+$ emission 
is significantly larger than the H$_2$O; however, deeper and higher resolution observations are needed
to better constrain these properties and to further interpret these results. \smallskip

\begin{table*}
\centering
\caption{The estimated size, based on a 2D fit to the moment-0 maps of each emission line. The size given corresponds to the FWHM of the emission, deconvolved from the beam.For W0149$+$2350 sizes were calculated for the restricted velocity range of $-300 < \upsilon < 750$\,km\,s$^{-1}$, to match the observed velocities of the truncated H$_2$O($2_{02}-1_{11}$) line. For W0410$-$0913, sizes were calculated for the velocity range of $-500 < \upsilon < 500$\,km\,s$^{-1}$, to match the observed velocities of the H$_2$O($3_{12}-3_{03}$) line. We note that the sizes do not change significantly if we do not apply these restrictions.}\label{tab:size}
\begin{tabular}{lccc}
\hline 
\hline 
Source & line & size & PA\\
&  & [arcsec$^2$] & [deg] \\
\hline
W0149$+$2350: & H$_2$O($2_{02}-1_{11}$) & $(0.30\pm0.08)\times(0.22\pm0.08) $ & $43\pm89$ \\ 
& CO(9$-$8) & $(0.43\pm0.07)\times(0.31\pm0.06) $ & $4\pm26$  \\ 
W0410$-$0913: & H$_2$O($3_{12}-3_{03}$) & $(0.41\pm0.08)\times(0.33\pm0.09)$ & $126\pm86$\\ 
& OH$^+$($1_{1}-0_{1}$) & $(1.3\pm0.3)\times(0.8\pm0.2)$ & $107\pm30$\\
& CO(9$-$8) & $(0.77\pm0.08)\times(0.43\pm0.05)$ & $108\pm8$\\ \hline
\end{tabular}	
\end{table*}

\noindent For both sources we find that the high-density gas traced by H$_2$O extends over a few kpc scales. Based on the total gas masses estimated by \cite{Fan18}, and if we assume a  spherical geometry\footnote{We note that if we instead assume a cylindrical geometry with a radius based on the estimated sizes from Table~\ref{tab:size} and a height of 0.5\,kpc, our results remain the same.} with radii of 1-2\,kpc and a volume filling factor of 0.01-0.1\%, then the estimated gas density is of the range of $n(H_2)\sim10^5-10^6$\,cm$^{-3}$. This is in agreement with what is required for the excitation of the observed H$_2$O emission. Therefore, the extent of the observed emission is consistent with the gas mass estimated for these galaxies.

\section{Discussion}

\subsection{Excitation of the H$_2$O and OH$^+$ emission} \label{sec:excitation}
In a detailed study of H$_2$O submillimetre lines in the nuclei 
of nearby star-forming galaxies, \cite{Liu17} used 
modelling to determine the physical conditions and excitation 
mechanisms needed for the H$_2$O transitions of low, 
medium, and high energy transition. \cite{Liu17} examined three 
components of the molecular ISM: a cold component with densities of $n(H_2)\sim10^4-10^6$\,cm$^{-3}$, 
dust temperatures of $T_d\sim20-30$\,K, and column densities of 
$N(H_2)\sim10^{23}$\,cm$^{-2}$; a warm  
component with $n(H_2)\sim10^5-10^6$\,cm$^{-3}$, 
$T_d\sim40-70$\,K, and 
$N(H_2)\sim1-4\times10^{24}$\,cm$^{-2}$; and a hot component with 
$n(H_2)\geq 10^6$\,cm$^{-3}$, $T_d\sim100-200$\,K, 
$N(H_2)\geq 5\times10^{24}$\,cm$^{-2}$. 
Interestingly, the H$_2$O($2_{02}-1_{11}$) transition can trace both the 
warm and hot components, but the H$_2$O($3_{12} - 3_{03}$) transition is stronger in 
the warm component of the ISM, and is not strongly produced in hot 
ISM conditions. Based on the radiative-transfer analysis of \cite{Liu17} 
the H$_2$O($2_{02}-1_{11}$) and H$_2$O($3_{12} - 3_{03}$) transitions can be produced by 
collisional excitation alone, while higher energy transitions 
characteristic of the hot ISM component require IR pumping to 
reach the observed intensities. However,  
\cite{Gonzalez-Alfonso14}, that also modelled the H$_2$O submillimetre lines
for IR galaxies, argue that IR pumping has a significant role in producing 
the H$_2$O($2_{02}-1_{11}$) and H$_2$O($3_{12} - 3_{03}$) line emission. 
In both cases, higher energy transitions are more likely to exclusively 
trace the hot component of the ISM surrounding AGN. Indeed, there was recently 
a detection of H$_2$O($4_{14}-3_{21}$) from the circumnuclear disk of a lensed quasar \citep{Stacey20}.

As we only have a single H$_2$O line detection for each source, it is not possible to make an extensive analysis. 
But we take a simplistic modelling approach using the non-LTE molecular radiative transfer code {\sc radex} \citep{vanderTak07}, to provide some constraints on the possible excitation of the observed H$_2$O emission.

We use the peak fluxes for each of the observed H$_2$O lines and the corresponding CO($9-8$) line, for each source. 
We assume a spherical symmetry, and CO column densities in the range of $10^{16}-10^{17}$\,cm$^{-2}$ to avoid 
optically thin emission at the low end and too high opacities at the high end.
Based on the SED fit and decomposition of these sources 
\citep[][]{Tsai15,Fan16b}, we expect an AGN 
torus dust temperature of 450\,K, and a dust temperature due to star-formation of 51\,K 
and 63\,K.

For W0149$+$2350 we find that the H$_2$O($2_{02}-1_{11}$) transition, at 987.94 GHz (E$_{\rm L}=$53\,K), may be collisionally 
excited. For intermediate densities ($\leq 10^6$\,cm$^{-3}$) relatively high H$_2$O($2_{02}-1_{11}$) abundances ($>10^{-6}$) are required. At higher densities ($> 10^6$\,cm$^{-3}$), lower abundances of $10^{-7}-10^{-8}$ are possible. 
For W0410$-$0913 we find that the H$_2$O($3_{12}-3_{03}$) transition, at 1097.36 GHz (E$_{\rm L}=$196\,K), may be collisionally excited, but that would require densities $> 10^6$\,cm$^{-3}$. Even for these high densities, the H$_2$O($3_{12}-3_{03}$) abundance is rather high ($10^{-5}-10^{-6}$). However, for both sources, having such high densities filling the region corresponding to the beam at these distances (i.e. $\sim$4--5\,kpc) is unlikely. This can either be solved with a small filling factor of dense clumps and/or the emission emerging from a dense nuclear structure (e.g. the AGN torus). Alternatively, all or parts of the H$_2$O emission are being influenced by IR pumping, which would allow it to arise also in lower density gas. This is in agreement with what is argued by \cite{Gonzalez-Alfonso14}.

Based on the results of our simplistic RADEX modelling and what has been found by \cite{Gonzalez-Alfonso14} and \cite{Liu17},
we expect that the transitions that we have detected are due to the high levels of star-formation in our 
sources, with a possibly significant contribution from the AGN (through IR pumping). 

So far in the literature, sources where the OH$^+$ line is in emission have been galaxies with significant 
AGN contributions such as Mrk231 \citep{vanderWerf10}, NGC1068 \citep{Spinoglio12}, and NGC7130 \citep{Pereira-Santaella13}. 
Therefore, the detection of OH$^+$ in emission for W0410$-$0913, 
could be indicative of excitation from cosmic rays and X-rays from the AGN 
\citep[e.g.][]{vanderWerf10,Spinoglio12,Pereira-Santaella13,Li20}. The relative line 
intensities between the CO($9-8$), the OH$^+$($1_1-0_1$) 
multiplet, and the H$_2$O($3_{12}-3_{03}$) observed for W0410$-$0913 are consistent with those of Mrk231 \citep{vanderWerf10}, 
for which it was shown that the combination of strong high$-J$ CO, and OH$^+$ emission indicate X-ray driven 
excitation from the AGN of Mrk231.
However, due to the limitations of our data, and the lack of other H$_n$O$^+$ transitions, we cannot confirm that 
the OH$^+$ emission is due to the AGN through radiative transfer modelling.

\subsection{The H$_2$O--IR luminosity ratio of Hot DOGs} \label{sec:lh2o_lir}

The relationship between the H$_2$O line luminosity with 
IR luminosity (from 8 to 1000\,$\mu m$; $L_{\rm IR}$), and the ratio of the two, has been well 
studied for low and high-$z$ dusty star-forming galaxies
 \citep[see  e.g.][]{Yang13,Yang16,Jarugula19}. Here we compare the two detected 
 Hot-DOGs of this paper to low and high-$z$ ultra-luminous IR galaxies (ULIRGs), sub-millimetre galaxies (SMGs) and AGN in the literature. We only 
 compare to sources with the same H$_2$O transition available, as each of the sources.

As our sources have both a strong AGN and star formation contribution to the total $L_{\rm IR}$, it is important to 
consider how we compare the H$_2$O line luminosity and the IR luminosity. For example, if the observed H$_2$O emission is 
primarily tracing the star formation in our source, then we should take the ratio with the IR luminosity due to star formation 
($L_{\rm IR, SF}$); however, if the emission is tracing both AGN and star formation regions then the total IR luminosity should 
be considered ($L_{\rm IR, tot}$). It is not possible in our analysis to confidently distinguish if the H$_2$O emission is only 
due to the star-formation of the source, as excitation from the AGN could also be contributing significantly.
For this reason we examine both of the possible scenarios when comparing to the literature.

\begin{figure}
\centering
\includegraphics[width=0.45\textwidth]{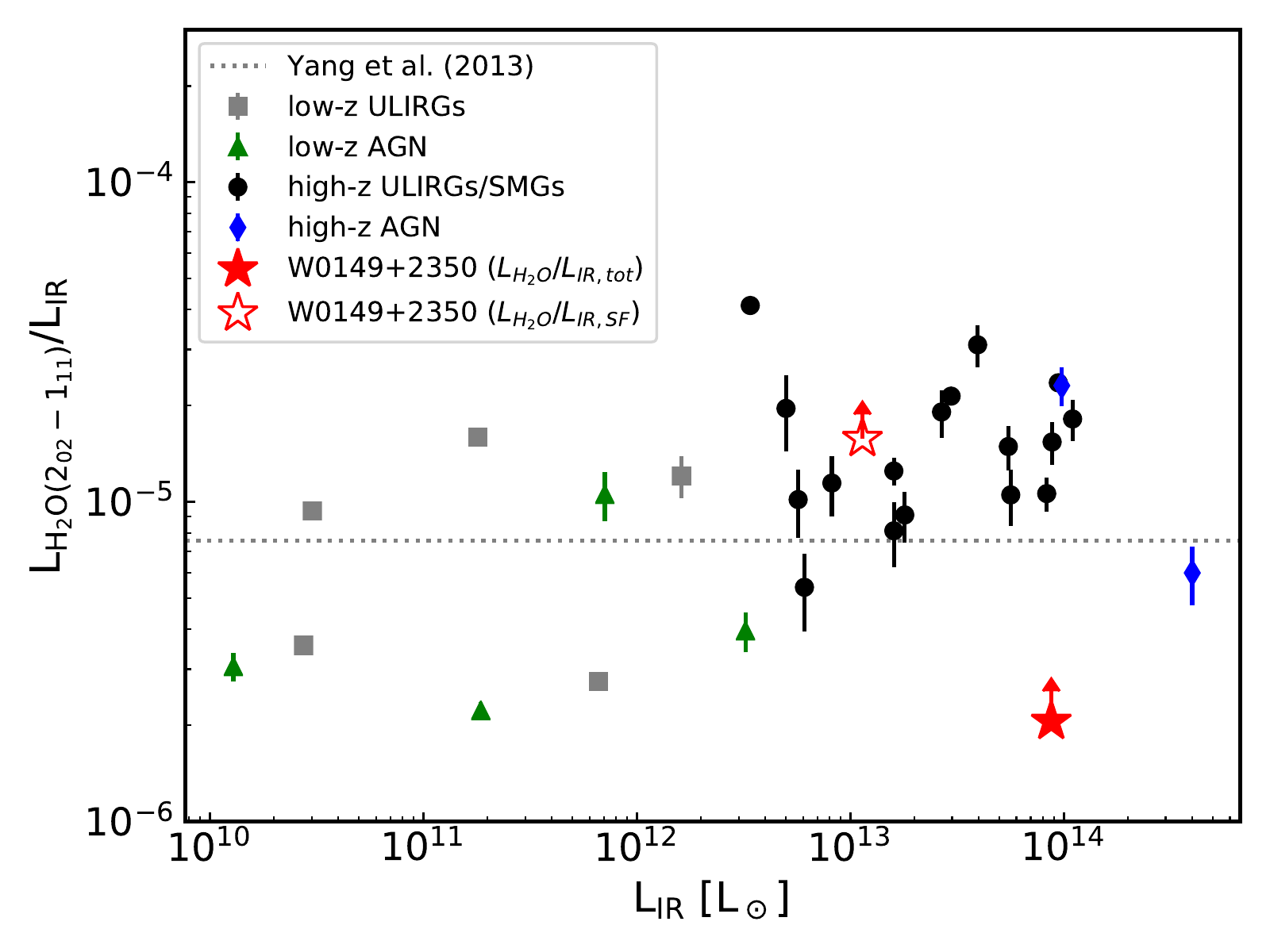} 
\includegraphics[width=0.45\textwidth]{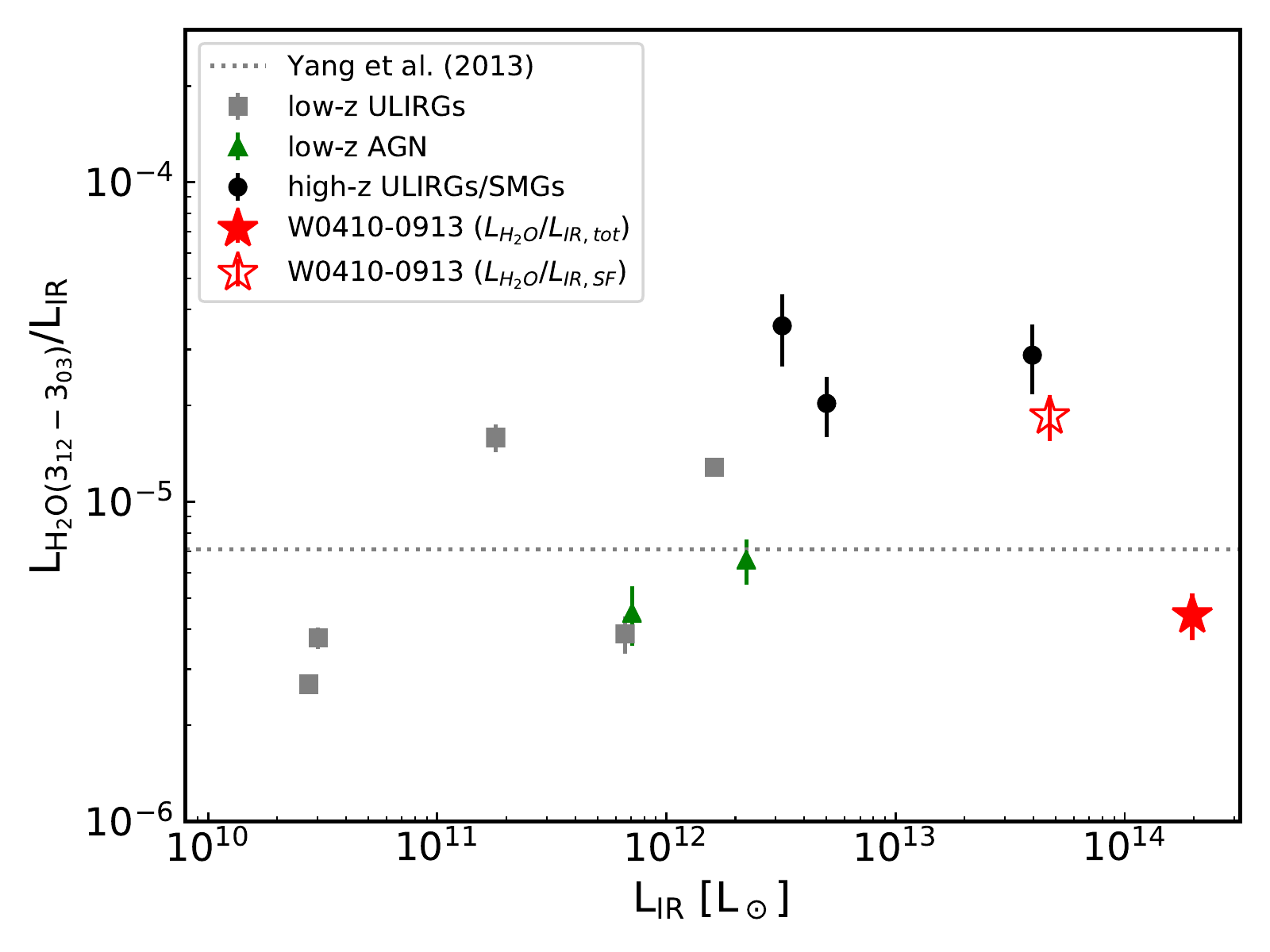} 
\caption{The ratio of the H$_2$O line luminosity over the IR luminosity ($L_{\rm H_2O}$/$L_{\rm IR}$), as a function of $L_{\rm IR}$, for the two Hot DOGs. {\em (top)} W0149$+$2350 in comparison to literature sources with detected $\rm H_2O(2_{02}-1_{11})$ emission, labelled based on their classification \citep{vanderWerf11,Combes12,Bothwell13,Omont13,Riechers13,Falstad15,Falstad17,Yang16,Liu17,Oteo17,Apostolovski19,Jarugula19}. Also shown are the average ratio found by \cite{Yang13} for nearby IR galaxies. {\em (bottom)} W0410$-$0913 in comparison to literature sources with detected $\rm H_2O(3_{12}-3_{03})$ emission, labelled based on their classification \citep{Riechers13,Falstad15,Falstad17,Liu17,Oteo17}, with the average ratio found by \cite{Yang13} for nearby IR galaxies also shown.}\label{fig:lir_lh2o}
\end{figure}

In Fig.~\ref{fig:lir_lh2o}{\em (top)} we show the ratio of the line luminosity for the $\rm H_2O(2_{02}-1_{11})$ transition over $L_{\rm IR}$ ($L_{\rm H_2O(2_{02}-1_{11})}$/$L_{\rm IR}$) as a function of $L_{\rm IR}$. Here we plot two values for W0149$+$2350, $L_{\rm H_2O(2_{02}-1_{11})}$/$L_{\rm IR, tot} > 0.25\times10^{-5}$ (solid star) and $L_{\rm H_2O(2_{02}-1_{11})}$/$L_{\rm IR,SF} > 2\times10^{-5}$ (hollow star), in comparison to both strong AGN and ULIRGs/SMGs that have been detected in the same transition. As we only have a lower limit for W0149$+$2350 the $L_{\rm H_2O(2_{02}-1_{11})}$/$L_{\rm IR, tot}$ ratio, although low, is consistent with both AGN and ULIRGs/SMGs at low and high-$z$. However, the lower limit for $L_{\rm H_2O(2_{02}-1_{11})}$/$L_{\rm IR, SF}$, is only consistent with luminous high-$z$ ULIRGs/SMGs, as can be expected if the source of the H$_2$O emission are the star formation regions.  

In Fig.~\ref{fig:lir_lh2o}{\em (bottom)} we show the ratio for the $\rm H_2O(3_{12}-3_{03})$ transition ($L_{\rm H_2O(3_{12}-3_{03})}$/$L_{\rm IR}$) as a function of $L_{\rm IR}$. Again, we plot two values corresponding to W0410$-$0913, $L_{\rm H_2O(3_{12}-3_{03})}$/$L_{\rm IR,tot} = 0.44\times10^{-5}$ (solid star) and $L_{\rm H_2O(3_{12}-3_{03})}$/$L_{\rm IR,SF} = 1.86\times10^{-5}$ (hollow star).
In this case the $L_{\rm H_2O(3_{12}-3_{03})}$/$L_{\rm IR,tot}$ ratio of W0410$-$0913 is consistent with low-$z$ AGN and ULIRGs, but is significantly below what is seen for high-$z$ ULIRGs/SMGs, while the $L_{\rm H_2O(3_{12}-3_{03})}$/$L_{\rm IR,SF}$ ratio is consistent with the high-$z$ ULIRGs/SMGs. 
 
\subsection{The H$_2$O and OH$^+$ line ratios of Hot DOGs}
In this section we compare the observed H$_2$O and OH$^+$ line ratios of our sources to the ratios in other galaxies found in the literature.

We find that for W0149$+$2350 the $H_2O(2_{02}-1_{11})$/CO(9$-$8) ratio is $>0.5$. This is in agreement with what is seen in the literature for high--$z$ ULIRGs/SMGs and AGN \citep[0.5--0.8 for ULIRGs/SMGs, and 0.4--0.7 for AGN; e.g.][]{Apostolovski19,Bradford09,Jarugula19,Li20,Oteo17,Riechers13,vanderWerf11,Weiss07}. However, the ratio of $H_2O(2_{02}-1_{11})$/CO(4$-$3) is $>1.4$, which is significantly higher than what has been observed for high--$z$ ULIRGs/SMGs \citep[0.4--0.85; e.g.][]{Bradford09,Jarugula19,Omont13,Yang16,Yang17}, although consistent with the luminous AGN APM08279$+$5255 that has a $H_2O(2_{02}-1_{11})$/CO(4$-$3) of 2.4 \citep{vanderWerf11,Weiss07}.

For W0410$-$0913 we calculate line ratios of $\rm H_2O(3_{12}-3_{03})/CO(9-8) = 0.8\pm0.2$, $\rm H_2O(3_{12}-3_{03})/CO(4-3) = 0.6\pm0.1$, and $\rm OH^+(1_{1}-0_{1})/H_2O(3_{12}-3_{03}) = 0.8\pm0.2$. In this case the available similar measurements in the  literature are limited. However, the $\rm H_2O(3_{12}-3_{03})/CO(9-8)$ ratio of W0410$-$0913 is consistent if somewhat higher than high--$z$ ULIRGs/SMGs \citep[0.4--0.75; e.g.][]{Oteo17,Riechers13}. The $\rm OH^+(1_{1}-0_{1})/H_2O(3_{12}-3_{03})$ ratio is consistent with low--$z$ luminous AGN \citep[e.g. Mrk231 and NGC1068;][]{vanderWerf10,Liu17,Spinoglio12}.

The nearby ULIRG Mrk231 is particularly interesting to compare to as it has a luminous AGN that contributes 70\% of the IR luminosity, similar to what is found for our Hot DOGs with 87\% and 77\% in W0149$+$2350 and W0410$-$0913, respectively \citep{Fan16b}. Mrk231 has $\rm H_2O(2_{02}-1_{11})$/CO(4$-$3) and $\rm H_2O(3_{12}-3_{03})/CO(4-3)$ line ratios of 0.4 \citep[e.g.][]{vanderWerf10,Papadopoulos07}, and an $\rm OH^+(1_{1}-0_{1})/H_2O(3_{12}-3_{03})$ ratio of 0.65 \citep{vanderWerf10,Liu17}. Overall, W0410$-$0913 seems most consistent with what is seen for Mrk231, while for W0149$+$2350 the $\rm H_2O(2_{02}-1_{11})$/CO(4$-$3) ratio is significantly larger than for Mrk231. Interestingly, W0149$+$2350 also has a significantly larger AGN contribution to the IR luminosity than W0410$-$0913 and Mrk231.

\section{Summary and Conclusions}
In this paper we have presented the recent detection of H$_2$O and OH$^+$ in $z>3$ Hot DOGs. 
This is the first detection of these emission lines for this class of galaxies, and one of the few in 
high-$z$ non-lensed galaxies \citep[e.g.][]{Casey19}. Specifically, we have detected the $\rm H_2O(2_{02}-1_{11})$ transition 
from W0149$+$2350, and the $\rm H_2O(3_{12}-3_{03})$, and the OH$^+$($1_{1}-0_{1}$) multiplet from W0410$-$0913.
These were serendipitous detections in an observation program targeting the CO(9$-$8) emission of these sources (Knudsen et al. in prep).

Similar to what has been found for previously observed high-$z$ 
ULIRGs and SMGs, the line profile of the H$_2$O emission lines follows that of the 
high-$J$ CO(9$-$8) emission. But the extent of the H$_2$O emission seems to be more 
compact than that of CO(9$-$8), and in the case of W0410$-$0913 it is more compact than the
OH$^+$($1_{1}-0_{1}$) emission too. However, deeper and higher resolution observations are needed
to better constrain the emission properties and to further interpret these results.

A single H$_2$O line detection does not allow for the 
disentangling of the ISM components and determining 
the molecular gas properties. However, the luminous 
H$_2$O emission in these two Hot DOGs 
indicate the existence of warm dense molecular gas conditions ($n(H_2)\sim10^5-10^6$\,cm$^{-3}$, 
$T_d\sim40-70$\,K),
possibly dominated by collisional excitation, with likely contribution from IR pumping from the AGN. 
We cannot confidently distinguish between excitation due to the 
AGN or star-formation or both for the H$_2$O emission. 
However, the detection of OH$^+$($1_{1}-0_{1}$) in emission for W0410$-$0913, and 
the agreement of the observed line ratios with luminous AGN in the literature indicate that 
the energy output from the AGN is dominating the radiative output of this galaxy, even though 
there is significant ongoing star formation (1000--5000\,M$_\odot$\,yr$^{-1}$). This is 
consistent with the fact that these galaxies host AGN that dominate the IR luminosity, 
contributing $\gtrsim70\%$ \citep{Fan16b}. This would also be consistent with the scenario 
that Hot DOGs are going through a transitional phase from a starburst--dominated to an AGN--dominated phase \citep[e.g.,][]{Eisenhardt12,Wu12,Fan16a}.

In order to break the degeneracies and disentangle the relative contributions from the AGN and the star formation, a multi-transitional approach is required. Based on 
the results of modelling studies \citep[e.g.][]{Gonzalez-Alfonso14,Liu17}, it is possible to design observational programs targeting a combination of H$_2$O transitions that trace different components of the molecular gas. Transitions such as $\rm H_2O(2_{02}-1_{11})$ and $\rm H_2O(3_{12}-3_{03})$, with energies of $E_{up} < 250-350$\,K, can be produced by collisional excitation alone or a combination of collisional excitation and IR pumping. However, higher energy transitions, such as H$_2$O($4_{14}-3_{21}$) or $\rm H_2O(4_{22}-4_{13})$, require IR pumping and are therefore more likely to directly trace excitation from the AGN. A combination of H$_2$O transitions from low, medium, and high energy levels, could be a useful tool for disentangling the AGN and star formation contributions to the excitation of the dense molecular gas.
Furthermore, targeting additional H$_n$O$^+$ transitions, such as H$_2$O$^+$ and H$_3$O$^+$ in combination with OH$^+$, would allow for constraints on the chemistry and excitation of the diffuse gas of these galaxies, possibly directly connected to the AGN \citep[see e.g.,][for detailed analysis]{Gonzalez-Alfonso13,Gonzalez-Alfonso18}.

\section*{Acknowledgements}
We thank the anonymous referee for constructive comments.  
We thank the staff of the Nordic ALMA Regional Center node for their support
and helpful discussions. We thank Dr Hannah Calcutt and Dr Pierre Cox for helpful discussions.
KK acknowledges support from the Knut and Alice Wallenberg Foundation. SA acknowledges support 
from the European Research Council (ERC) and the Swedish Research Council. LF acknowledges the support from the National Natural Science Foundation of China (NSFC Nos. 11822303, 11773020 and 11421303) and Shandong Provincial Natural Science Foundation (JQ201801).
This paper makes use of the following ALMA data:
ADS/JAO.ALMA\#2017.1.00123.S. ALMA is a partnership of ESO (representing
its member states), NSF (USA) and NINS (Japan), together with NRC
(Canada) and NSC and ASIAA (Taiwan) and KASI (Republic of Korea), in 
cooperation with the Republic of Chile. The Joint ALMA Observatory is 
operated by ESO, AUI/NRAO and NAOJ.

\bibliographystyle{aa}
\bibliography{waterHDs.bib}

\end{document}